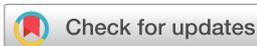

# Critical effect of the bottom electrode on the ferroelectricity of epitaxial Hf$_{0.5}$Zr$_{0.5}$O$_2$ thin films†

Saúl Estandía, [a] Jaume Gàzquez, [a] María Varela, [b] Nico Dix, [a] Mengdi Qian, [a] Raúl Solanas, [a] Ignasi Fina [a] and Florencio Sánchez [*a]

Epitaxial orthorhombic Hf$_{0.5}$Zr$_{0.5}$O$_2$ (HZO) films on La$_{0.67}$Sr$_{0.33}$MnO$_3$ (LSMO) electrodes show robust ferroelectricity with high polarization, endurance and retention. However, no similar results have been achieved using other perovskite electrodes so far. Here, LSMO and other perovskite electrodes are compared. A small amount of orthorhombic phase and low polarization are found in HZO films grown on La-doped BaSnO$_3$ and Nb-doped SrTiO$_3$, while null amounts of orthorhombic phase and polarization are detected in films on LaNiO$_3$ and SrRuO$_3$. The critical effect of the electrode on the stabilized phases is not a consequence of the differences in the electrode lattice parameter. The interface is critical, and engineering the HZO bottom interface on just a few monolayers of LSMO permits the stabilization of the orthorhombic phase. Furthermore, while the specific divalent ion (Sr or Ca) in the manganite is not relevant, reducing the La content causes a severe reduction of the amount of orthorhombic phase and the ferroelectric polarization in the HZO film.

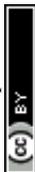

## 1. Introduction

The discovery of ferroelectricity in a metastable phase of doped HfO$_2$[1] has attracted great interest due to its potential impact on memories and other devices. Most of the research on ferroelectric HfO$_2$ has been conducted with polycrystalline films, generally grown by atomic layer deposition. Epitaxial films, with a more controllable microstructure, may allow a better understanding of ferroelectric properties and prototyping devices. Examples of the usefulness of doped HfO$_2$ epitaxial films include the achievement of coercive field ($E_c$)–thickness ($t$) $E_c \sim t^{-2/3}$ scaling,[2,3] the separation of ionic and electronic transport contributions in ferroelectric tunnel junctions,[4] or the achievement of simultaneous high polarization, endurance and retention in films thinner than 5 nm.[5]

Epitaxial films are being investigated now very actively. Several conducting or semiconducting substrates or templates, including Si wafers,[6] TiN,[7] GaN,[8] and indium tin oxide,[9] have been used to deposit epitaxial films of ferroelectric doped HfO$_2$. In contrast, epitaxy of hafnia on perovskite-type oxide electrodes has only been reported on Nb-doped SrTiO$_3$ (Nb:STO)[10] and La$_{0.67}$Sr$_{0.33}$MnO$_3$ (LSMO).[2,11–17] Epitaxy on other popular perovskite electrodes, such as SrRuO$_3$ (SRO) or LaNiO$_3$ (LNO), has not been achieved as far as we know. This is in stark contrast to conventional ferroelectric perovskites, which can be grown epitaxially on a large number of metal and oxide electrodes.[18–20] It should be noted that the epitaxy of ferroelectric hafnia is challenging because it requires the formation of a metastable orthorhombic phase instead of the paraelectric phase that is stable in bulk. In epitaxial growth, a low interface energy is critical for the stabilization of metastable phases.[17,21] Therefore, for epitaxial stabilization of ferroelectric hafnia, the matching of the lower electrode with the orthorhombic phase has to be better than with competing polymorphs. This restricts the number of suitable electrodes for epitaxial stabilization.

We have carried out a systematic study of the impact of the lower electrode on the epitaxial stabilization of the ferroelectric hafnia. A series of Hf$_{0.5}$Zr$_{0.5}$O$_2$ (HZO) films were deposited on a set of conducting epitaxial oxide thin films: LNO, LSMO, SRO and Ba$_{0.95}$La$_{0.05}$SnO$_3$ (BLSO) grown on STO(001). A doped Nb:STO substrate was also used as the electrode. X-ray diffraction (XRD) measurements confirmed the epitaxial stabilization on LSMO of the orthorhombic (o) ferroelectric phase, (111)-oriented, and accompanied by a minority monoclinic phase, in agreement with previously reported results.[2,11,16] The o-phase was also detected on the HZO film directly grown on the Nb:STO substrate, being the (111)-crystallites tilted by around 15°. Similar tilted epitaxy was observed on BLSO, the o-HZO(111) crystallites being accompanied in this case by monoclinic (−111) crystallites and tilted by around 19°. The o-phase was not detected in the films on LNO and SRO. In agreement with the XRD analysis of the films, the measured ferroelectric polarization was high in HZO films on LSMO, low in films on Nb:STO and BLSO, and null in films on LNO and SRO. To determine whether the critical impact of LSMO on the stabilization of the ferroelectric phase is mainly caused by epitaxial stress or whether the interface chemistry is critical, we also used lower electrodes formed by two stacked conductive oxides, the one interfacing HZO being around only three unit cells thick. The obtained results rule out the epitaxial stress as the main factor making the electrode critical. Finally, we have also explored the use of manganite electrodes with different chemical compositions (LaMnO$_3$ doped with either Sr or Ca) and with different amounts of doping (half-doped or optimally doped). The content of La in the manganite is found to determine the amount of orthorhombic phase present along with ferroelectric polarization, while the use of either Sr or Ca has no effect.

[a] Institut de Ciència de Materials de Barcelona (ICMAB-CSIC), Campus UAB, Bellaterra 08193, Spain. E-mail: fsanchez@icmab.es
[b] Dpto. Física de Materiales & Instituto Pluridisciplinar, Universidad Complutense de Madrid, 28040, Madrid, Spain

† Electronic supplementary information (ESI) available: XRD θ–2θ scans measured with a point detector. Scanning transmission electron microscopy (cross-sectional Z-contrast images) of the HZO films on bilayer electrodes with ultra-thin SRO and LNO layers in contact with the HZO. See DOI: 10.1039/d0tc05853j

## 2. Experimental

The perovskite electrode and the HZO film were grown in a single process by pulsed laser deposition (KrF excimer laser) on SrTiO$_3$(001) substrates. HZO films, 9.5 nm thick, were deposited at 800 °C (substrate heater block temperature) under 0.1 mbar of molecular oxygen, and they were cooled after growth under 0.2 mbar of oxygen. A set of conducting perovskite thin films were used as the electrode: LaNiO$_3$, four La$_{1-x}$A$_x$MnO$_3$ (A = Sr, Ca; $x$ = 0.33, 0.5) manganites, SrRuO$_3$, and Ba$_{0.95}$La$_{0.05}$SnO$_3$. Substrate temperature and oxygen pressure during deposition of the electrodes were 700 °C and 0.15 mbar for LaNiO$_3$, 700 °C and 0.1 mbar for La$_{1-x}$A$_x$MnO$_3$, 700 °C and 0.2 mbar for SrRuO$_3$, and 725 °C and 0.1 mbar for Ba$_{0.95}$La$_{0.05}$SnO$_3$, respectively. An HZO film was also deposited on a conducting 0.5% mol Nb-doped SrTiO$_3$(001) substrate. Additional information on growth conditions and structural properties of the electrodes is reported elsewhere.[22–24]

The crystal structure was characterized by XRD with Cu Kα radiation using a Siemens D5000 diffractometer with a point detector and a Bruker D8, equipped with a two-dimensional detector Vantec 500. A sketch of the angles with respect to the surface normal can be seen in Fig. S1 (ESI†). Top platinum electrodes, 20 nm in thickness and around 20 μm in diameter, were deposited by DC magnetron sputtering through stencil masks. Ferroelectric polarization loops of the capacitors were measured at room temperature in top-bottom configuration by means of an aixACCT TF Analyser 2000 platform. Leakage contribution was compensated for using the positive up negative down (PUND) procedure.[25]

## 3. Results and discussion

In the first series, HZO films were deposited either on STO(001) substrates buffered with conducting LNO, LSMO, SRO and BLSO layers (Fig. 1a) or on a bare conducting Nb:STO(001) substrate. The lattice parameter of the electrodes in bulk (Fig. 1b)

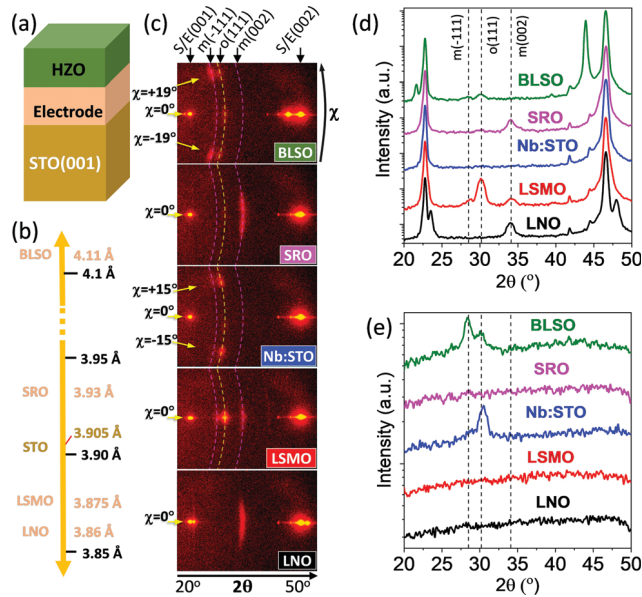

Fig. 1 (a) Sketch of the epitaxial HZO/electrode heterostructures on STO(001). (b) Lattice parameters (bulk) of the perovskite electrodes. (c) XRD 2θ–χ frames. The electrode is indicated at the bottom right of each frame. Dashed yellow and pink lines mark the 2θ positions of the observed orthorhombic and monoclinic reflections, respectively. These reflections, as well as the (001) and (002) reflections of the substrate (S) and the electrode (E), are marked also by arrows in the top of the panel. The 2θ range is indicated in the frame shown at the bottom. (d) θ–2θ scans obtained by integrating χ in the −10° to +10° range or a smaller range with the aim of maximizing the signal/background ratio. (e) Equivalent θ–2θ scans obtained by integrating χ in the −22° to −12° and +12° to +22° ranges or smaller ranges to maximize the signal/background ratio.

varies within a wide range from 3.86 Å (LNO) to 4.11 Å (BLSO), while the lattice parameter of the STO substrate is 3.905 Å. Fig. 1c shows the XRD 2θ–χ frames (2θ from 20° to 50°, χ from around −20° to +20° for the smallest 2θ values) of the five samples. The highest intensity spots in all samples, located at χ = 0° (corresponding to the surface normal) and 2θ around 22.8° and 46.5°, correspond to (001) and (002) reflections of substrates and electrodes, respectively. The other spots are reflections from the HZO film. The film on LSMO exhibits a very bright spot at χ = 0° and 2θ ∼ 30.2°, corresponding to the o-HZO(111) reflection. There is also a weaker intensity spot, centered at χ = 0° and elongated from around χ = −5° to +5°, at 2θ ∼ 34.1°, corresponding to the position of the monoclinic (002) reflection. Thus, the HZO film on LSMO presents a (111)-oriented orthorhombic phase along with a minority monoclinic (001)-oriented phase. This finding can be appreciated also in the θ–2θ scan obtained by integration around χ = 0° (Fig. 1d, red line). A detailed structural characterization of equivalent samples confirmed that both phases are epitaxial.[2,11,16] In contrast, the films on LNO and SRO do not show the reflection from the orthorhombic phase, but they also exhibit the elongated monoclinic (002) spot at 2θ ∼ 34.1°. The integrated scans are shown in Fig. 1d (black and pink lines for LNO and SRO, respectively). On the other hand, the frames of the HZO films on Nb:STO and BLSO do not display evident symmetric reflections (χ = 0°) from the film. The integrated scans around $\chi = 0°$ in Fig. 1d show no HZO peaks for the film on Nb:STO (blue line) and a tiny o-HZO(111) peak for the film on BLSO (green line). However, the XRD frames of the two films exhibit asymmetrical reflections that did not appear in the other samples. In the film on Nb:STO, there is a pair of spots at $2\theta$ around 30.2°, corresponding to o-HZO(111), and $\chi = -15°$ and $+15°$. This finding indicates tilted epitaxy,[26] with o-HZO(111) crystallites being tilted by ~15° with respect to the normal direction. A similar pair of tilted o-HZO(111) spots is present in the film on BLSO, but this time at $\chi = -19/+19°$ and having a weaker intensity. Moreover, these spots are accompanied by another couple of spots, at the same $\chi = -19/+19°$ and $2\theta \sim 28.5°$ that corresponds to monoclinic HZO(−111). The $\theta$–$2\theta$ scans obtained by integration around $\chi = -19°$ (from −22° to −16°) and +19° (from +16° to +22°) (Fig. 1e) evidence the presence of these asymmetrical reflections in the films on Nb:STO (blue line) and BLSO (green line), and its absence in the other samples.

PUND measurements were conducted to evaluate the ferroelectricity of the HZO films (Fig. 2). The current–voltage (I–V) curves (Fig. 2a) and the corresponding polarization loops (Fig. 2b) confirm the expected ferroelectricity of the film on LSMO, with remanent polarization $P_r \sim 10~\mu\text{C cm}^{-2}$ and coercive electric field $\sim 3.4~\text{MV cm}^{-1}$. Magnified I–V curves and polarization loops of the other samples are depicted in Fig. S2, ESI.† The films on Nb:STO (blue curve) and BLSO (green curve) show ferroelectric switching peaks at around 4 V (Fig. 2a and Fig. S2a, ESI†), signaling the ferroelectric nature of the samples. Note that the loops in Fig. 2b are not saturated, so accurate polarization values cannot be given. The film on SRO shows a hysteretic I–V curve, but without a peak. This result is in agreement with a minor I–V loop (where $V_{\text{max}} < V_c$) of a ferroelectric material. However, the I–V curves collected at higher voltages result in device failure before observing any current peak. The absence of a ferroelectric current switching peak and the absence of orthorhombic phase traces in the sample (Fig. 1) indicate that the switchable current results from non-ferroelectric contributions, most probably intrinsic electroresistance. The I–V curve of the film on LNO is similar to the one on SRO, but with significantly lower current values. The ferroelectric measurements are thus in agreement with the distinct presence of the orthorhombic phase in the samples (Fig. 1), confirming the critical effect of the LSMO electrode on the epitaxial stabilization of the orthorhombic phase. Note that the shape of the loop and the integrated polarization change for the LSMO, SRO, Nb:STO and BLSO samples during the first cycles (Fig. S3, ESI†) and the plotted data correspond to the loops collected after 10 cycles.

Epitaxial stress has been reported to have a major influence on the stabilization of the orthorhombic phase on LSMO electrodes.[16] However, the notorious differences in stabilized phases and the orientation of the HZO films on the set of five electrodes do not correlate with the bulk lattice parameter of the electrodes (Fig. 1b). Nevertheless, the electrode thin films could be strained, and, to estimate accurately the out-of-plane lattice constants of the thin film electrodes, $\theta$–$2\theta$ scans around symmetrical reflections were acquired with a point detector (Fig. S4, ESI†). The out-of-plane lattice parameters, determined from the positions of the corresponding (002) diffraction peaks, are $d_{\text{LNO}} = 3.79$ Å (3.86 Å in bulk), $d_{\text{LSMO}} = 3.86$ Å (3.875 Å in bulk), $d_{\text{SRO}} = 3.94$ Å, the value extracted from the fit of two Gaussian curves to STO(002) and SRO(002) (3.93 Å in bulk), and $d_{\text{BLSO}} = 4.12$ Å (4.1 Å in bulk), indicating that LNO, LSMO and SRO are strained, and BLSO is plastically relaxed. The in-plane lattice parameter of LNO, LSMO and SRO is expected to be coincident with the lattice parameter of the STO substrate (as measured in ref. 16 and 27), very similar to that of the conducting Nb:STO(001) substrate, and significantly smaller than that of relaxed BLSO. Therefore, a completely different epitaxial stabilization in this series of films takes place despite the almost coincident in-plane lattice parameter (and thus the lattice mismatch with HZO polymorphs) of four of the samples in the series.

Since the critical impact of the electrodes on the epitaxial stabilization of the ferroelectric phase is not caused by differences in the epitaxial stress, other factors that can modify the interface energy and therefore the epitaxial stabilization need to be considered. To evaluate whether the different chemical compositions and atomic structures of the surface of the electrodes are relevant factors, HZO films were deposited on SRO and LNO electrodes covered by an ultra-thin (~1.2 nm) LSMO layer and on LSMO electrodes covered by an ultra-thin (~1.2 nm) LNO or SRO layer (Fig. 3a). The XRD $2\theta$–$\chi$ frames (Fig. 3b) demonstrate that the effect of the ultra-thin layer interfacing HZO is huge. While HZO

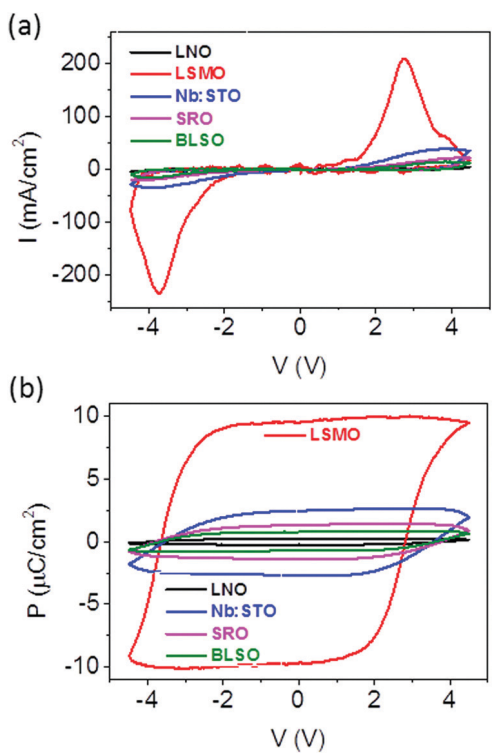

Fig. 2 (a) I–V curves of the HZO films grown on different perovskite electrodes and (b) the corresponding polarization loops.

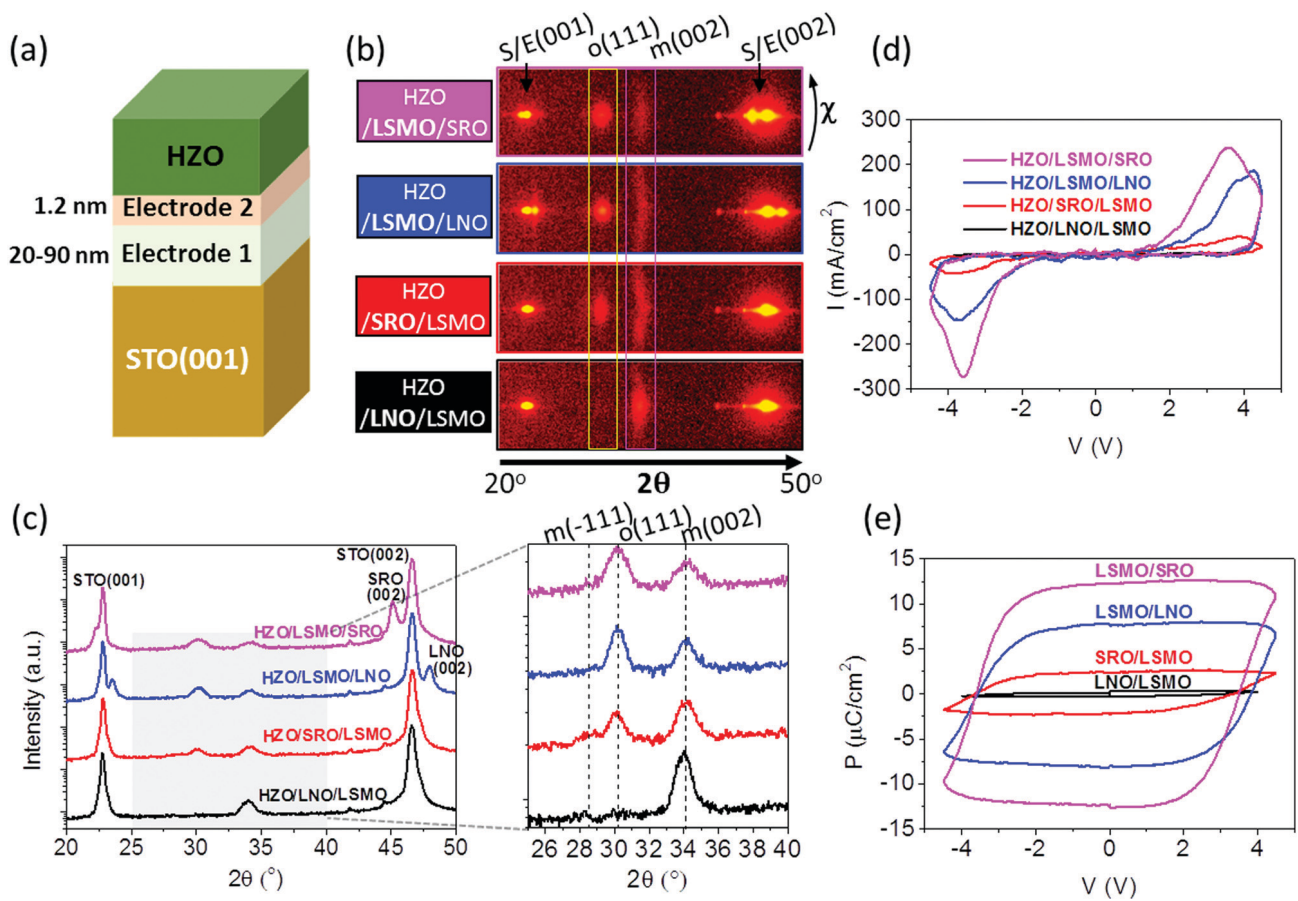

Fig. 3 (a) Sketch of the epitaxial HZO/bilayer electrode heterostructures, in which the electrode interfacing HZO is 1.2 nm thick. (b) XRD $2\theta-\chi$ frames of the samples. The bilayer electrode is indicated at the left of each frame. Vertical yellow and pink rectangles mark the orthorhombic o-HZO(111) and monoclinic m-(002) reflections, respectively. The (001) and (002) reflections of the substrate (S) and electrodes (E) are marked by arrows. (c) $\theta-2\theta$ scans obtained by integrating the XRD $2\theta-\chi$ frames in the $\chi$ −10° to +10° range. Right: Zoomed scans around the monoclinic m-HZO(002) and m-HZO(−111), and orthorhombic o-HZO(111) reflections. (d) $I-V$ curves and (e) the corresponding polarization loops of the HZO films.

films grown on bare SRO or LNO are purely monoclinic, the films on either LSMO/SRO or LSMO/LNO are mostly orthorhombic, being in fact the frames very similar to that of the HZO film grown on bare LSMO. In contrast, the o-HZO(111) reflection is very weak in the film on SRO/LSMO, and in the film on LNO/LSMO there is only a high intensity spot corresponding to the monoclinic (002) reflection. The integrated $\theta-2\theta$ scans in Fig. 3c summarize the effect of the 1.2 nm thick layers on the stabilization of the different HZO polymorphs. The $\theta-2\theta$ scans measured with a point detector show equivalent results (Fig. S5, ESI†). The $I-V$ curves (Fig. 3d) and the corresponding polarization loops (Fig. 3e) confirm the huge effect: polarization is high when the ultra-thin layer interfacing HZO is LSMO, very small when this layer is SRO, and null when the layer is LNO. It is important to note that the non-negligible amount of orthorhombic phase present in the film on SRO/LSMO originates from an incomplete coverage of the LSMO surface by SRO,[22] as shown by cross-sectional scanning transmission electron microscopy images of the interface (Fig. S6, ESI†).

HZO has to grow on LSMO for the orthorhombic phase to stabilize, and, as reported in ref. 16, the amount of orthorhombic phase depends on the strain of the LSMO when different substrates are used. Probably, the chemical composition and crystal structure of the LSMO surface reduce the energy at the interface between the electrode and the orthorhombic phase, favoring its formation instead of the other HZO polymorphs with a lower energy in bulk. To gain an insight about this phenomenon, a series of four HZO/manganite/STO(001) samples (Fig. 4a) were prepared using manganite electrodes with different chemical compositions, $La_{1-x}A_xMnO_3$ (A = Sr, Ca; $x$ = 0.33, 0.5). The XRD $2\theta-\chi$ frame scans in Fig. 4b and the integrated $\theta-2\theta$ scans in Fig. 4c show that there are no significant differences between the two films grown on optimally doped manganites ($x$ = 0.33): a bright o-HZO(111) and a low intensity elongated monoclinic (002) spot, indicating that the epitaxial orthorhombic phase constitutes the majority for films on both Sr and Ca optimally doped manganites. In contrast, the frames of the two films on half-doped manganites ($x$ = 0.5) show weaker o-HZO(111) diffraction spots, particularly on the Ca-doped electrode, whereas the intensity of the elongated monoclinic (002) spot is increased in both films (equivalent results are obtained for the $\theta-2\theta$ scans measured with a point detector, Fig. S7, ESI†). The $I-V$ curves and polarization loops (Fig. 4d and e) confirm that the

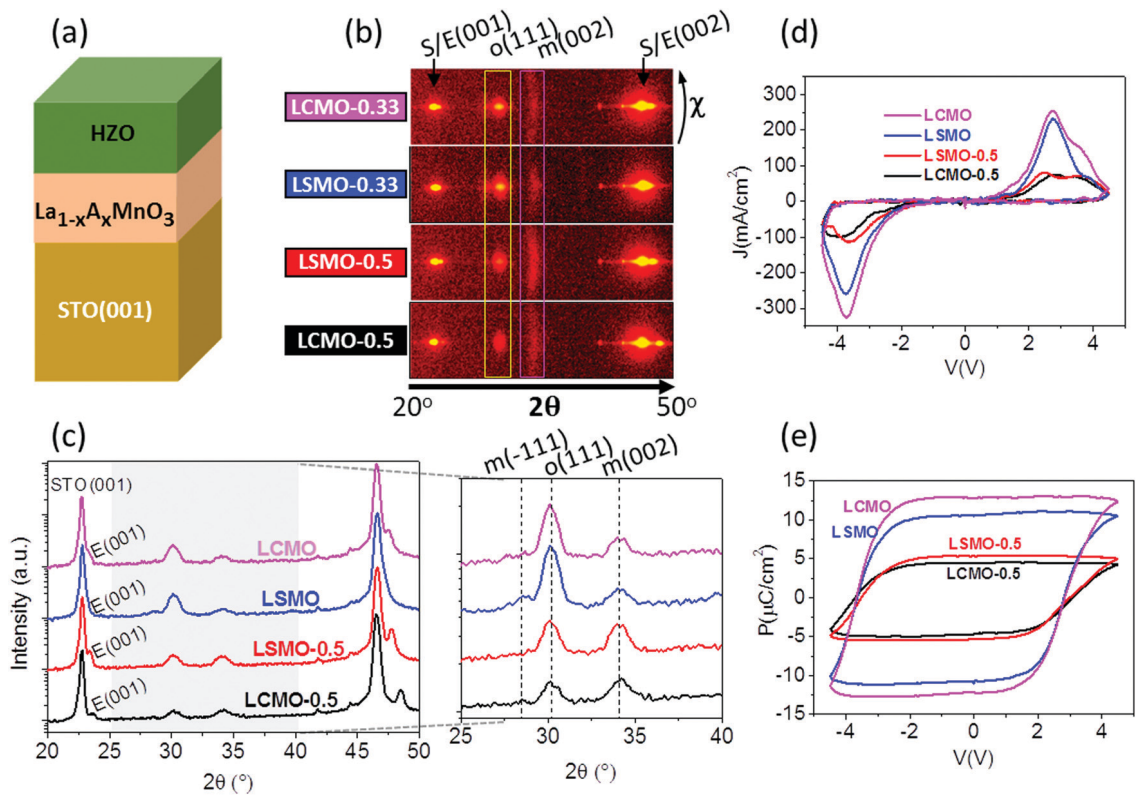

Fig. 4 (a) Sketch of the epitaxial HZO/La$_{1-x}$A$_x$MnO$_3$ heterostructures. (b) XRD 2$\theta$–$\chi$ frames of the samples. The La$_{1-x}$A$_x$MnO$_3$ composition (A = Sr or Ca, and $x$ = 0.33 or 0.5) is indicated at the left of each frame. Vertical yellow and pink rectangles mark the orthorhombic o-HZO(111) and monoclinic m-(002) reflections, respectively. The (001) and (002) reflections of the substrate (S) and electrodes (E) are marked by arrows. (c) $\theta$–2$\theta$ scans obtained by integration in $\chi$ between −10° and +10°. Right: Zoomed scans around the monoclinic m-HZO(002) and orthorhombic o-HZO(111) reflections. (d) I–V curves and (e) the corresponding polarization loops of the HZO films.

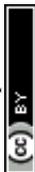

four films are ferroelectric. The remanent polarization on both optimally doped electrodes ($x$ = 0.33) is more than 10 μC cm$^{-2}$, whereas the films on half-doped manganites ($x$ = 0.5) have much lower polarization (Fig. 4e), in agreement with the smaller amount of orthorhombic phase observed. Small differences have been observed during the first cycles as shown in Fig. S8 (ESI†).

The results shown in Fig. 4 prove that epitaxial stabilization of the ferroelectric HZO phase is reduced in films on manganite electrodes with a lower La content. This finding can be due to a dependence on the La content of the interface energy between the manganite and orthorhombic HZO. Differences in the electrical conductivity of the manganites could also be a factor to be considered. In bulk, half-doped manganites have much lower room-temperature conductivity than the corresponding optimally doped compounds. The electrical resistivity values of elastically strained La$_{0.5}$Sr$_{0.5}$MnO$_3$ and La$_{0.5}$Ca$_{0.5}$MnO$_3$ epitaxial films on STO(001) are around 10$^{-3}$ and 1 Ω cm, respectively.[23] The resistivity of the former is thus low enough to be used as an electrode, while the higher resistivity of the latter could produce a depolarization field. However, the ferroelectric polarization of HZO on these electrodes is similar, despite the large difference in the electrode resistivity, thus disregarding the possibly relevant role of the electrostatic boundary conditions in the formation of the ferroelectric HZO phase. Therefore, the higher polarization of the films grown on optimally doped electrodes is a result of the greater amount of orthorhombic phase, which is likely due to an enhanced epitaxial stabilization on manganite electrodes with a larger La content, regardless of whether the divalent atom is Sr or Ca.

Overall, our study confirms that the interface chemistry influences critically the stabilization of the metastable orthorhombic phase and consequently the ferroelectric polarization in epitaxial HZO films. In the case of ferroelectric perovskites, the effects of the interface chemistry on the polarization are known. In particular, the chemical termination of the LSMO bottom electrode can have a huge effect on the polarization, as demonstrated in BiFeO$_3$[28] and PbTiO$_3$[29] films. The orthorhombic phase of HZO forms on manganite electrodes, and its formation is favored on La$_{1-x}$A$_x$MnO$_3$ with a greater La content, while HZO on LNO is monoclinic despite the presence of La atoms. La$_{1-x}$A$_x$MnO$_3$ and LNO are grown on STO substrates, which present a majority of TiO$_2$-termination, and therefore La$_{1-x}$A$_x$MnO$_3$ and LNO are expected to be mostly MnO$_2$ and NiO$_2$ terminated, respectively. Thus, the impact of La content in the manganite electrode on the epitaxial stabilization of the orthorhombic phase points to a complex interface. Indeed, the scanning transmission electron microscopy characterization revealed an intriguing interface, with a pseudomorphic HZO single layer and a semicoherent HZO film above.[17] Other heteroepitaxial oxide systems are stabilized also by lattice mismatch accommodation through changes in the interface chemistry. For example, the formation of a $(Ti,V)O_2$ coherent interphase layer allows the epitaxial stabilization of the metastable monoclinic $VO_2(B)$ phase on STO substrates.[30] Detailed studies of the atomic structure at the HZO/LSMO interface may allow the understanding of the intriguing mechanisms of the epitaxial stabilization of the orthorhombic phase.

## 4. Conclusions

In conclusion, manganite electrodes are critical to epitaxially stabilize the ferroelectric orthorhombic phase of hafnia. The films deposited on other investigated conducting perovskite electrodes present low (on Nb:$SrTiO_3$ and $Ba_{0.95}La_{0.05}SnO_3$) or null (on $LaNiO_3$ and $SrRuO_3$) orthorhombic phase and ferroelectric polarization. The need of manganites to stabilize the HZO ferroelectric phase is not related to epitaxial stress. We show that engineering the interface with a few LSMO monolayers permits stabilizing the ferroelectric phase. Yet, while the effectivity of the $La_{1-x}A_xMnO_3$ manganites does not depend on the divalent ion A, being that either Sr or Ca, it is found to be greater for optimally doped manganites ($x = 0.33$) than for manganites with a lower La content.

## Conflicts of interest

There are no conflicts to declare.

## Acknowledgements

The financial support from the Spanish Ministry of Science and Innovation through the Severo Ochoa FUNFUTURE (CEX2019-000917-S), MAT2017-85232-R (AEI/FEDER, EU), and PID2019-107727RB-I00 (AEI/FEDER, EU) projects and from Generalitat de Catalunya (2017 SGR 1377) is acknowledged. IF and JG acknowledge Ramón y Cajal contracts RYC-2017-22531 and RYC-2012-11709, respectively. Project supported by a 2020 Leonardo Grant for Researchers and Cultural Creators, BBVA Foundation and by CSIC through the i-LINK program (LINKA20338). SE acknowledges the Spanish Ministry of Economy, Competitiveness and Universities for his PhD contract (SEV-2015-0496-16-3) and its cofunding by the ESF. SE's work has been done as a part of his PhD program in Materials Science at Universitat Autònoma de Barcelona. The electron microscopy observations carried out at the Centro Nacional de Microscopía Electrónica at UCM (MV) were supported by MICINN grant# RTI2018-097895-B-C43. We acknowledge support of the publication fee by the CSIC Open Access Publication Support Initiative through its Unit of Information Resources for Research (URICI).

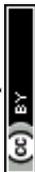